%% file: trustsense.tex
\documentclass[conference,10pt]{IEEEtran}
\usepackage{graphicx}
\usepackage{array}
\usepackage{calc}
\usepackage{algorithm}
\usepackage{algorithmic}
\usepackage{amsmath}
\usepackage{subfigure}

\begin{document}

	\title{TrustSense: An energy efficient trust scheme for clustered wireless sensor networks }

	\author{\IEEEauthorblockN{Adedayo Odesile}
		\IEEEauthorblockA{Department of Computing and Software Systems\\
			University of Washington Bothell\\
			Email: adedao@uw.edu}
		\and
		\IEEEauthorblockN{Brent Lagesse}
		\IEEEauthorblockA{Department of Computing and Software Systems\\
			University of Washington Bothell\\
			Email: lagesse@uw.edu}
	}
		
	\maketitle
	\begin{abstract}
	Designing security systems for wireless sensor networks presents a challenge due to their relatively low computational resources. This has rendered many traditional defense mechanisms  based on cryptography infeasible for deployment on such networks. Reputation and anomaly detection systems have been implemented as viable alternatives, but existing implementations still struggle with providing efficient security without a significant impact on energy consumption. To address this trade-off between resource consumption and resiliency, we designed TrustSense, a reputation management protocol for clustered WSNs. It is a semi-centralized family of algorithms that combine periodic trust updates, spatial correlation, and packet sequence validation at the cluster-heads' hierarchy to relieve the  sensor nodes of unnecessary opinion queries and trust evaluation computation. We compared the efficiency of TrustSense with legacy reputation systems such as EigenTrust and the results of simulations show a significant improvement in reliability and energy usage while maintaining an acceptable path length with varying numbers of malicious nodes. We believe the approach of combining different techniques from various classes of intrusion detection systems unlocks several possibilities of achieving better results by more complex and versatile composition of these techniques. 
	\end{abstract}
	\input{introduction}

	\input{related_work}
	\input{design}

	\input{simulations}

	\input{conclusion}
	\bibliographystyle{IEEEtran}
	\bibliography{trustsense}

\end{document}

%% file: introduction.tex
\section{Introduction}

Wireless Sensor Networks (WSNs) are comprised of spatially distributed sensor nodes, observing certain phenomena and cooperatively routing measured information to one or more central locations. WSNs have application in a variety of fields such as facility management, smart-homes, traffic controls, incidence response, and other diverse areas. In general, contemporary ubiquitous systems involving the Internet of Things and pervasive computing include aspects of WSNs, so they would also benefit from our research. Unlike traditional networks, WSNs are characterized by energy constrained nodes, dynamism in topology, and are mostly ad-hoc in nature. As such, they are considered be easier and less expensive to deploy. However, these advantages come at a security cost, increasing the chances of an attacker gaining access and eventually subverting the network through various malicious activities. This is mostly the case in unattended and non-tamper-proof networks where including a malicious node or re-engineering an existing benevolent node is a real possibility. Given the wide range applications of WSNs, the importance of establishing an efficient means of protecting the network from adversarial threats cannot be over-emphasized. As a result, numerous research studies\cite{survey} on securing these systems have been carried out in the past few decades.
There have been two broad categories of security mechanisms researched and implemented for WSNs which are the cryptographic and non-cryptographic techniques \cite{evidential}. Cryptographic implementations focus more on intrusion prevention through various authentication protocols mostly involving the use of digital signatures. If the adversary eventually by-passes them and gains control of a single node in the network, all the other nodes are considered vulnerable. Secondly, the computational overhead created by cryptographic implementations are unsuitable for WSNs in most cases due to their resource constraint.

Conversely, non-cryptographic techniques like reputation and anomaly detection systems are more concerned with isolating adversaries that are actively participating in the network. Current security research in WSNs \cite{penalty,btrm,eigen,tms,nees} is geared towards this approach as the attacker is engaged throughout the process, constantly being faced with the challenge of survival in the network before eventually being isolated. Focusing on intrusion detection and isolation as opposed to prevention also minimizes the impact it has on the openness and flexibility of WSNs. Depending on the implementation, this category of systems could also create high computation and communication overhead. This brings a challenge of balancing the trade-offs between resiliency and energy consumption in these intrusion detection/isolation systems which remains an active research problem.

In this paper, we developed a trust management scheme "TrustSense" that leverages novel ideas from some existing implementations \cite{nees,tms} to provide a rich and energy-efficient system that achieves a better balance between resiliency and energy trade-off. It combines the use of periodic trust-updates and caching for reducing communication overhead, linear spatial variogram for outlier detection and a strict reward system to quicken the isolation of malicious nodes. Based on our simulation test results, we were able to simultaneously achieve resiliency despite a high percentage of malicious nodes in the network while consuming less energy than previous approaches.

%% file: related_work.tex
\section{Related Works}
An extensive survey on detection and isolation of malicious behaviour in WSNs was carried out by Illiano et al. \cite{survey} where they classified these systems into anomaly detection and trust based systems respectively. The anomaly detection systems compared reported values from sensors with an expected data model and ﬂag a node as suspect/malicious if the difference between measured and expected value exceeds a certain threshold. These data models were built on different correlation techniques. They were further classified by Illiano et al. into four namely: 
\begin{itemize}
	\item \textbf{Temporal Correlation} which compares the currently measured value of a single node with a series of its recently measured values to detect abnormalities.
	
	\item  \textbf{Spatial Correlation} whereby measurements sampled at any point in time from a group of sensors observing the same thing within a common location are examined and outliers are ﬂagged.
	
	\item \textbf{Physical Attribute Correlation} where change in measured values are compared with the constraints that characterize the physical phenomena being observed to detect abnormal/unrealistic changes. For example, it is impossible to measure a temperature value of -1 Kelvin and as such any sensor reporting that will be considered an outlier.
	
	\item \textbf{Hybrid Systems} that combine two or more of the above techniques.
\end{itemize}

\cite{spatial_wu,spatial_ngai} implemented spatial correlation models under the assumptions of spatial homogeneity which suggests that similar sensors within a certain area should measure the same values with no variation except when it random errors occur. Zhang et al. relaxes this constraint by modeling spatial correlation as a function of the distance between each sensor \cite{relaxed_spatial}, thereby allowing more variations. We focused on the spatial correlation model because of the challenges posed by the other two. Temporal correlation models make it difficult to discern between actual incidence spikes and malicious activities because both cases will result into violation of the correlation constraint. The physical attribute correlation is context specific and cannot be generalized to every WSN.

Trust-based systems measure the confidence level that the behavior of a certain node is acceptable. While all trust systems exhibit similar high level function in that they acquire information, compute trust, and disseminate that trust\cite{lagesse_dtt:_2009}, the way trust values are computed, assigned and accessed differentiates existing implementations. Gomez et al. \cite{btrm} implemented a trust algorithm, BTRM, that models the behavior of ants in a colony whereby the ants leave pheromone traces along a navigated path to help subsequent ants find their way to a certain destination. The ants in this context refer to the control packets sent by reputable nodes that modify existing pheromones they encounter on their way from source to destination node. The pheromones are modified by a function that helps other sensors identify the most trustworthy (with high probability) path to send data. 

Abramov and his peers implemented TMS \cite{tms}; a trust algorithm based on their previous trust model in \cite{tms_model}. The trust model encompasses metrics that characterize different malicious behaviors. Local trust of a node B from the perspective of node A is computed as a weighted normalized difference between the number of successful and failed interactions A has had with B for every situation corresponding to a related metric. Once A decides that B is misbehaving, it reports B to the cluster head which in turn makes a final decision about B. They compared their algorithm with the BTRM, EigenTrust and, PowerTrust, realizing a lower resource usage and high resiliency against most of their tested threat models except the Sybil attacks. Hoceini et al. took a slightly different approach in their NEES trust model implementation \cite{nees} which was focused on conserving energy by reducing communication overhead. They used periodic trust updates as opposed to other trust models that compute trust values for every request. However, their algorithm only addresses malicious packet drops. To counter malicious nodes that perform on-off attacks in which they behave benignly for longer periods of time to build strong reputations and occasionally perform malicious acts, Oh et al. \cite{penalty} decided to assign a penalty to reduce the tolerance that the system has for malicious behavior, thereby speeding up the isolation of misbehaving nodes.

It was observed that each of the existing trust systems were designed based on diverse concepts that addressed different security issues. In TrustSense, we merged some of these concepts to generate a solution that strikes an optimal balance between efficiency and immunity. 

%% file: design.tex
\section{System Design}

The decisions that guided the design of TrustSense are explained in this section. The first design consideration was the threat model which was scoped as follows:

\subsection{Threat Model}
TrustSense was designed to protect WSNs against adversaries actively carrying out malicious activities in the network that could range from sink-hole attacks to data falsification while minimizing its energy footprint. Dealing with passive listeners illegally extracting information from the network is outside the scope of non-cryptographic mechanisms, and as such was not considered in the design.

\subsection{TrustSense Design Decisions}

In order to improve upon existing trust management systems, we identified recurring questions in every reputation model and addressed them in the following subsections.

Aggressive conservation of energy was a primary consideration in arriving at the subsequent design decisions which were to favor caching over computation, computation over communication, and in order to reduce predictability of the system from malicious nodes, heavy use of randomization was also infused into the protocol.

\subsubsection{Types of topologies system could work with}
As stated in the title, TrustSense was designed to work strictly with a clustered topology. This is due to the heavy reliability on cluster-heads to facilitate the trust protocol.

\subsubsection{Assumptions that could be made about the nodes in the network}
The communication between a node and its neighbors are assumed to be reliable, hence, this will not work for intermittent high-latency, opportunistic or very sparse networks. Secondly, the cluster-heads are assumed to be higher power tamper-proof nodes with longer range of communication such that they could reach out to other cluster-heads and all the sensors within their respective clusters via one-way multi-casts.

\subsubsection{Prior information needed}
In order to facilitate coordination and control by the cluster-heads, the following list of parameters need to preloaded: trust update period, maximum outlier thresholds, median and high reputation value thresholds, packet relay and consistency reward rate, packet loss and outlier punishment rate, and spatial relevance.  

\subsubsection{How new nodes get reputation information}
\begin{figure}[ht!]
		\centering
		\includegraphics[width=85mm]{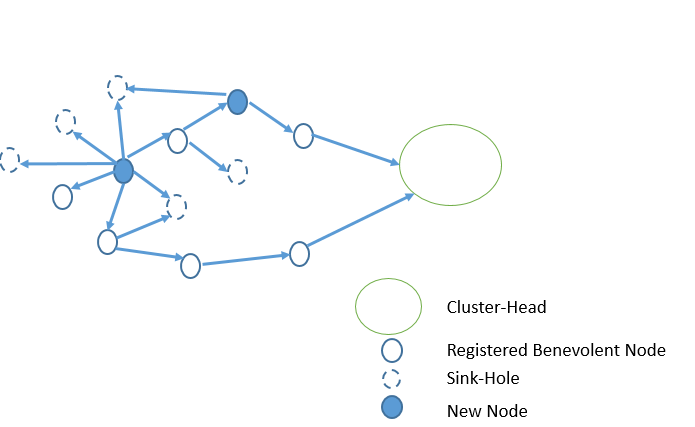}
		\caption {Directional Flood Protocol}
		\label{fig:flood}
	\end{figure}

On deployment of nodes into the network, and subsequent inclusion of new nodes, a pseudo-random Id is generated for each of them. Every node is responsible for getting its registration packet to the nearest cluster-head. Only then is the node considered to be part of the network or else the cluster-heads will drop any data reported from it. Since there is a possibility of the new nodes being surrounded mostly by attackers, a directional flooding protocol is used to ensure that the registration packet gets delivered to a cluster-head as shown in figure \ref{fig:flood}.

New nodes have no idea of the closest cluster-head's location, hence they broadcast registration packet to all their neighbors. Registered benevolent neighbors will examine the packet for location information and forward only if they are closer to the cluster-head than the new node. The only caveat is if the new node is totally surrounded by sink-holes, it is ineluctably denied from joining the network. 

On reception of the registration packet, the cluster-head adds new node's information to its local-cluster table and it is subsequently assigned a default trust value interpolated between the minimum good and medium threshold as: 
	\[default trust=0.25 * gd + 0.75 * md\label{equ:one}\tag{1}\].
However, if the node was coming from another known cluster, its Id will be detected in the cluster-head's global trust table, hence its trust value will be assigned from the appropriate entry. To aid with spatial correlation computation, a location graph encoded as a vector of coordinates in each entry on the local trust table is updated if newly added node is close enough to be a neighbor of the node corresponding to each entry. The node information will automatically be included in the next local and global update. Algorithm \ref{alg:newnodereg} demonstrates how the registration process occurs.
\begin{algorithm}
	\caption{New Node Registration}
	\label{alg:newnodereg}
	\begin{algorithmic}
		\REQUIRE $lclTableCount < maxCount$
		\IF {$lclTable.contains(regPacket.Id)$}
		\STATE return
		\ENDIF
		\STATE $defTrust \leftarrow 0.25 * goodThrsh + 0.75 * mediumThrsh$
		\IF {$glbTable.contains(regPacket.Id)$}
		\STATE $defTrust\leftarrow glbTable.get (regPacket.Id).trust$
		\ELSE
		\STATE $glbTable.add(GlobalEntry(regPacket.Id,defTrust))$
		\ENDIF
		\STATE $newEntry \leftarrow lclTrust.add(LocalEntry(regPacket.Id,$
		\STATE $defTrust,regPacket.location,regPacket.IPLink))$
		\FORALL {$e\ IN\ lclTable$}
		\IF {$e.Id \neq newEntry.Id$}
		\IF {$Abs(e.location-newEntry.location)<maxRange$}
		\STATE $e.spatialNeighbors.add(newEntry.Id)$
		\STATE $newEntry.spatialNeighbors.add(e.Id)$
		\ENDIF
		\ENDIF
		\ENDFOR 
	\end{algorithmic}
\end{algorithm}

\subsubsection{ How trust/reputation evaluated}
 TrustSense uses periodic trust updates.  As opposed to constantly querying neighbors about a potential forwarding nodes, getting reputation information is only a matter of doing a simple look-up on your local cache which is guaranteed to be consistent with every other node's opinion. This is because the cached trust information is all coming from one source; the cluster-head.

For fair distribution of data forwarding to give newer nodes a chance to build their reputation, a pseudo-random  path selection seed is also attached to the trust updates and is constantly regenerated in subsequent updates. This seed is passed into a link selection function described in Algorithm \ref{alg:linkselection}. The seed is used to randomly select a trust range which could either depict a highly trusted, trusted or suspicious node with the probability distribution skewed to the highly trusted nodes as shown in algorithm \ref{alg:linkselection}.
\begin{algorithm}
	\caption{Link Selection Function}
	\label{alg:linkselection}
	\begin{algorithmic}
		\REQUIRE $seed \leq 100\ AND\ \geq 0 $
		\STATE $minTrust \leftarrow 0$
		\STATE $maxTrust \leftarrow 1.0$
		\IF {$seed < 60$}
		\STATE $minTrust \leftarrow goodThrs$
		\ELSIF {$seed > 60\ AND\ seed < 90$}
		\STATE $minTrust \leftarrow medThrs$
		\STATE $maxTrust \leftarrow goodThrs$
		\ELSE
		\STATE $maxTrust \leftarrow medThrs$
		\STATE $minTrust \leftarrow double.minvalue$
		\ENDIF
		\STATE $candidateNodes \leftarrow empty$
		\REPEAT 
		\FORALL {$n\ IN\ neighbors.distancemap$}
		\IF  {$n.distancetoCh \leq this.distanceToCh$}
		\STATE $currTrust \leftarrow defaultTrust$
		\IF {$trustTable.get(n.id) \neq NULL$}
		\STATE $currTrust \leftarrow trustTable.get(n.id)$ 
		\ENDIF
		\IF {$currTrust \geq minTrust\ AND\ currTrust \leq maxTrust$}
		\STATE $candidateNodes.add(n)$
		\ENDIF
		\ENDIF
		\ENDFOR
	
		\IF {$candidateNodes.empty$}
		\IF {$minTrust \equiv goodThrs$}
		\STATE $minTrust \leftarrow medThrs$
		\STATE $maxTrust \leftarrow goodThrs$
		\ELSIF {$minTrust \equiv medThrs$}
		\STATE $minTrust \leftarrow double.minValue$ 
		\STATE $maxTrust \leftarrow medThrs$
		\ELSE
		\STATE $maxTrust \leftarrow double.minvalue$
		\ENDIF
		\ENDIF
		\UNTIL {$candidateNodes.count > 0\ OR\ maxTrust \equiv double.minvalue$}
		\IF {$candidateNodes.count > 0$}
		\STATE $size \leftarrow candidateNodes.size $
		\STATE $forwardingNode \leftarrow candidateNodes.get$
		\STATE $(seed\ MOD\ size)$
		\ENDIF
	\end{algorithmic}
\end{algorithm}

\subsubsection{Representation of trust}
We represented trust as any real number ranging from 0 to 1. We then chose 0.25 as the minimum trust value a node must have to be considered trusted (medium  threshold), and 0.75 for highly trusted nodes. Any node below 0.25 is considered a primary suspect. Once the trust value falls below 0, node is black-listed from the network.  These values can be modified, but they produced good results in simulation.

\subsubsection{Trust management/distribution}
Trust information is maintained by cluster-heads in two different tables. A local table containing a rich set of details about nodes within its cluster and a global table containing an Id to reputation mapping of both intra and inter cluster sensor nodes within the network. Every trust update period, a global update occurs where all cluster heads exchange a minified version of their local trust tables containing just Ids and trust values, and a list of nodes ids they have black-listed in their local clusters in order for other cluster-heads to remove them from their global trust tables. 

Subsequently, a local update is triggered where every cluster-head multi-casts relevant information from their local trust table to nodes registered within their respective cluster. Each node sieves out only information concerning its neighbors and updates its local trust cache and distance map if needed. The newly updated black-list is also sent along with the updates so nodes are aware of the sensors they should avoid.

\subsubsection{How reputation is computed}
Trust computation is carried out by cluster-heads. To reduce computational complexity, we use simple increments and decrements for rewards and punishment respectively.  This approach has been shown to be effective in prior work \cite{lagesse_arex:_2008}. The values to be added or deducted in both cases are part of the initialization parameters embedded in the cluster-head. However, these values are not always fixed, especially in cases where reward/punishment is spread amongst multiple nodes.

\subsubsection{Classification of activities}
Good behaviors consist of a) reporting accurate data values, b) consistently delivering packets to the cluster heads with minimal loss, and c) forwarding packets on behalf of other sensors. On the other hand, a) reporting suspicious data values, b) delivering packets out of order, and c) selective forwarding or total route denial are considered punishable.

\subsubsection{Identification of good behavior and isolation/punishment of bad ones}
Based on the previous subsection, the list of good and bad behaviors are seen to be just one-to-one mapping of contrasting activities; meaning the same algorithm holds for identifying a good behavior and its direct bad opposite.

\textbf{Starting with data accuracy}, a spatial correlation technique based on a linear variogram as demonstrated by Zhang et al. in \cite{relaxed_spatial} was used. Using a variogram relaxes the constraint of every sensor within a spatial domain having to report the same values. An expected value was computed as a weighted average of values acquired from one-hop neighbors and the actual reported data value from the sensor, with the weights being a linear function of the distance between neighboring node and the actual sensor whose value is being examined. If the difference between expected value and the actual value is greater than a certain threshold, data is flagged as outlying. This computation is carried out by the cluster head before data bundling for the base-station. Below is the equation that models the expected value E(x) for an arbitrary sensor x:
		\[W_{i}=\frac{d(x,i)}{d_{max}}\label{equ:two}\tag{2}\]
		\[E(x)= \frac{(\sum_{i}^{n} W_{i}* V(i)) + V(x)}{(\sum_{i}^{n} W_{i}) +1} \label{equ:three}\tag{3}\]
In equation \ref{equ:two}, the spatial weight is computed using the linear variogram function where \emph{d(x,i)} represents the distance between sensor x and neighbor i, and $d_{max}$ is the maximum one-hop distance between two sensors. In equation \ref{equ:three}, \emph{E(x)} is the expected value for sensor x and \emph{V(x)} is the actual value reported by sensor x which is infused into the variogram as the most influential value in computing \emph{E(x)}.

If \emph{E(x)} and \emph{V(x)} are consistent, the node x is rewarded with 50\% probability by incrementing its trust value by the consistency reward rate. On the other hand, the node has its outlier counter \emph{O(x)} incremented in the local trust table. If \emph{O(x)} becomes equal to the maximum event outlier sequence, it indicates that something else must be wrong around the node's spatial domain for the consistent outlying values which could be a concentrated incidence spike or a defect in the node. Such incidents are reported to the base station to resolve. However, if the node reports a consistent value before the  maximum outlier sequence is reached, it depicts (with high probability) an intentional on-off malicious data injection and node is punished immediately by having its trust value decremented in the trust table. Its \emph{O(x)} value is also reset to 0.

\textbf{For packet forwarding and delivery}, every data sample originating from an arbitrary node is tagged with its node Id and a data sequence. On random occasions, with roughly 50\% probability, the cluster-head carries out a packet loss/drop inspection. It checks for the chronology of a registered node x's reported data packets by comparing the measured data sequence \emph{S(x)} and with the one currently cached in the local-trust table \emph{S($x_{cached}$)}. If \emph{S(x)} - \emph{S($x_{cached}$)} is not equal to one, it means a data packet was skipped hence, the cluster-head decides to punish suspect nodes. However, nodes are rewarded for forwarding with a 50\% probability if sequence numbers were consistent. Using the last generated path selection seed, cluster-head estimates the most probable path missing/reported packet must have travelled with the link selection function described in algorithm II and randomly distributes punishments/rewards across nodes included in the path. This list is not exclusive of the originating node. 

\textbf{Lastly, in keeping the local-trust table up to date}, cluster-head reduces the spatial presence value of any node that missed a data bundling window. If node reports any value before hitting a spatial presence of 0, it is reset back to 1, or else the node is removed from both the local and global trust table. 

The reputation model was tested through simulations using the TRMSim-WSN v0.5 developed by Abramov et al. \cite{simulator}.

%% file: simulations.tex
\section {Experimental Simulations and Results}
We conducted experiments using two broad scenarios. The first one is such that the malicious nodes stays the same throughout the run, and the second in which the nodes oscillates between malicious and benevolent status at random while still maintaining the percentage of malicious nodes in the network.  Under each scenario, we increased the percentage of malicious nodes in 10\% steps and recorded the average satisfaction, path to reliable servers and relative energy consumption for 100 sensors executing 30 times per run which is the minimal number of samples needed to approximate a normal distribution. In all cases, TrustSense was compared with existing trust models implemented in the simulator which includes BTRM \cite{btrm}, EigenTrust \cite{eigen}, PowerTrust \cite{powertrust} and PeerTrust \cite{peertrust}.

Figure 3 and 4 show the energy consumption measured as a function of transmitted packet distance for both stable and oscillating adversaries. Interestingly, there is no visible relationship between the percentage of malicious nodes and the energy consumed in all five trust models. EigenTrust's energy usage was seen to exceed the rest in order of magnitude. This is expected as it was designed mainly for traditional P2P systems and not WSNs. The other trust models consumed similar amount of energy except for PeerTrust which was significantly lower with extreme percentages of malicious nodes while peaking at approximately 50\% threat level. This is because the algorithm is based on certainty functions that require more computation to distinguish malicious and benevolent nodes when they are fairly equal in number. TrustSense consumed the least amount of energy in most of the threat level cases due to reduced communication overhead using periodic trust updates and caching of neighbor reputation information.

\begin{figure*}[t]
\centering
\subfigure[Average Energy Usage Plot]{
    \includegraphics[width=0.3\textwidth]{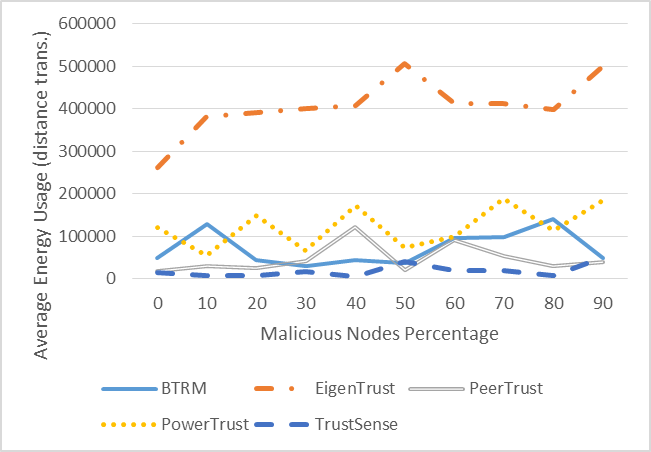}
    \label{fig:energy}
}
\subfigure[Average Accuracy Plot]{
    \includegraphics[width=0.3\textwidth]{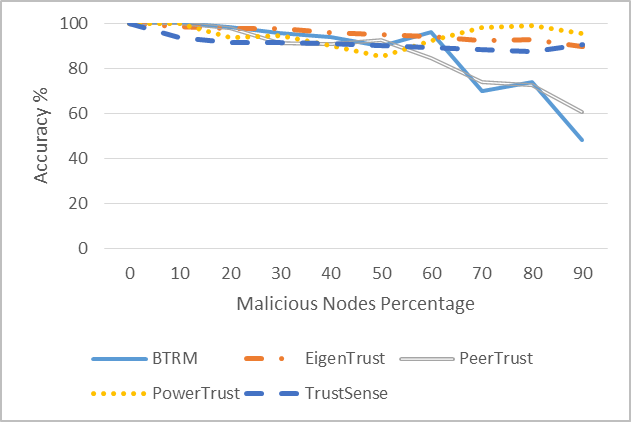}
    \label{fig:acc}
}
\subfigure[Average Path Length Plot]{
    \includegraphics[width=0.3\textwidth]{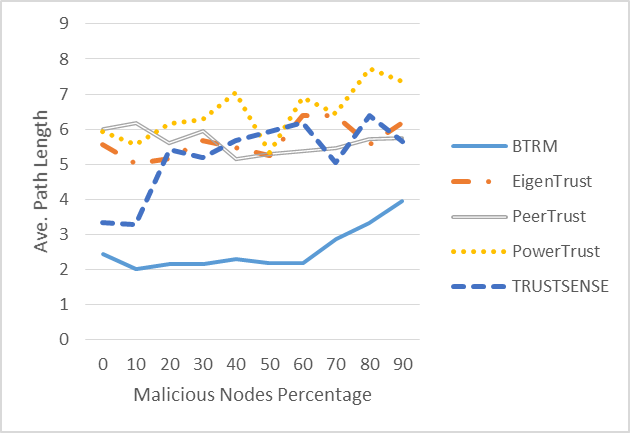}
    \label{fig:len}
}
\caption[Optional caption for list of figures]{Static Performance}
\label{fig:staticperformance}
\end{figure*}

\begin{figure*}[t]
\centering
\subfigure[Average Energy Usage Plot]{
    \includegraphics[width=0.3\textwidth]{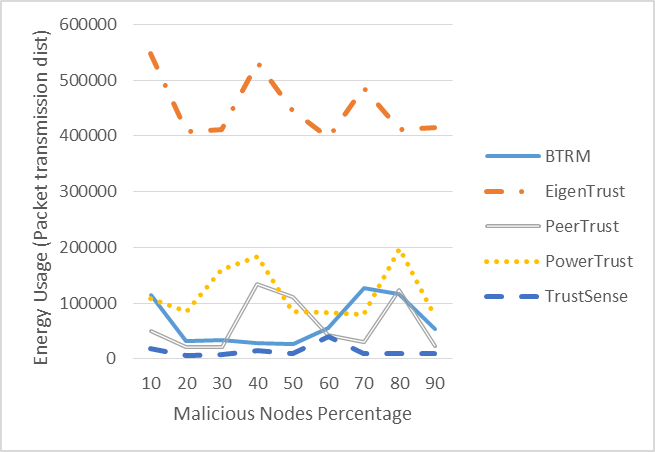}
    \label{fig:oscenergy}
}
\subfigure[Average Accuracy Plot]{
    \includegraphics[width=0.3\textwidth]{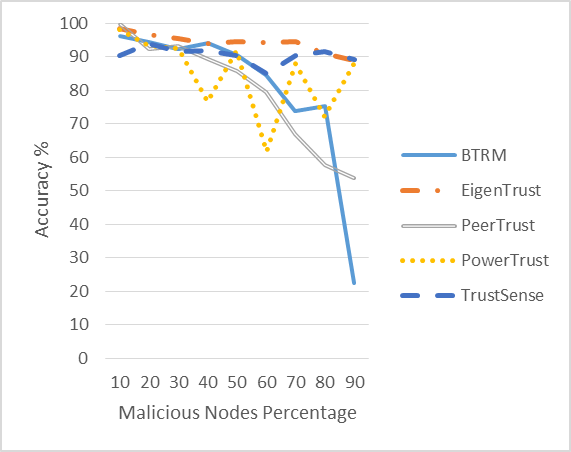}
    \label{fig:oscacc}
}
\subfigure[Average Path Length Plot]{
    \includegraphics[width=0.3\textwidth]{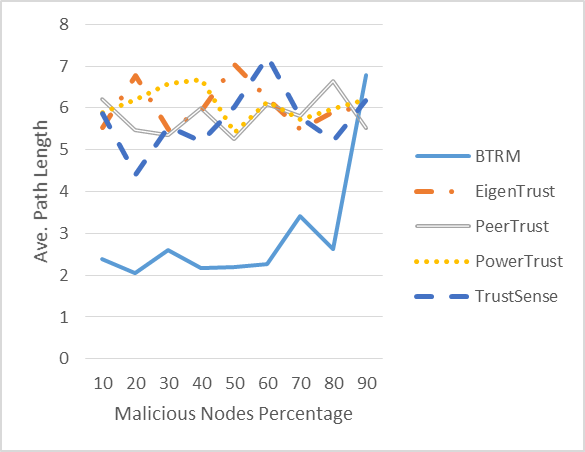}
    \label{fig:osclen}
}
\caption[Optional caption for list of figures]{Oscillating Performance}
\label{fig:oscperformance}
\end{figure*}

Figure 5 and 6 show the accuracy plot which represents the probability of getting a satisfactory service/delivery. BTRM and PeerTrust were seen to drop significantly in this regard once the threat level crosses 60\% while the network infiltrated with 90\% of malicious nodes. As expected, worse results were realized in the oscillating adversary mode for all trust models. EigenTrust happened to be the most stable and decreased steadily with increasing threats while PowerTrust fluctuated but maintained an approximately equal level of accuracy with EigenTrust. TrustSense maintained a competitive value just below EigenTrust.

The average number of hops taken to reach a reliable cluster-head/server is shown in figure 7 and 8. BTRM had the fewest hop counts while the others took over twice the path length to successfully deliver packets. Except for PeerTrust, other trust models showed a fairly positive correlation between the threat level and the average path length. TrustSense was on par with the majority of other trust models due to randomness in its link selection entry algorithm that may not always pick the closest node to the cluster-head. 

After examining all three metrics and their relative importance in determining a quantitative estimation of the energy-security trade-off, we normalized and assigned respective weights to them in order to arrive at a percentage raw score.

	\[A_{N}=\frac{A}{A_{max}}\label{equ:four}\tag{4}\]
	\[P_{N}=1-\frac{P}{P_{max}}\label{equ:five}\tag{5}\]
	\[E_{N}=1-\frac{E}{E_{max}}\label{equ:six}\tag{6}\]
	\[Score = (W_{A} * A_{N}) + (W_{P} * P_{N})  +  (W_{E} * E_{N})\label{equ:seven}\tag{7}\]
	
Where $A$ = accuracy, $P$ = average path length, $E$ = average energy usage, and $_{N}$ is the normalized flag. $A_{max}$ was set to 100\%, $P_{max}$ was set to the total number of sensors in the network, basing the value on an assumption that a packet transverses the whole network in worst case scenarios, and $E_{max}$ was set to the nearest 100,000 from the maximum recorded value because there is nothing within the simulation limiting/capping energy consumption. The values 40, 20 and 40 were assigned to the weights $W_{A}$, $W_{P}$, and $W_{E}$ respectively to add up to a maximum trade-off score of 100. The average path length was considered least important because it is more strongly related to latency than security or energy.

\begin{figure}[ht!]
	\centering
	\includegraphics[width=85mm]{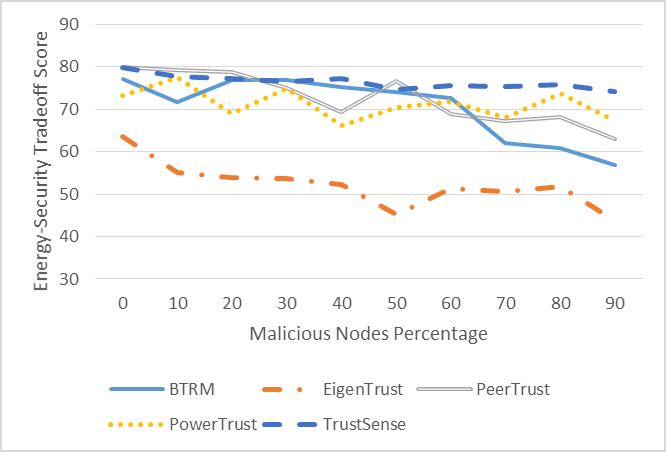}
	\caption {Energy-Security Trade-off Score Plot}
	\label{fig:tradeoff}
\end{figure}

\begin{figure}[ht!]
	\centering
	\includegraphics[width=85mm]{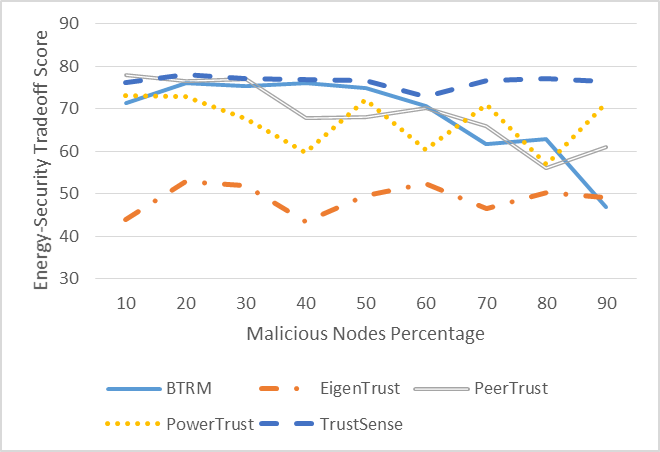}
	\caption {Energy-Security Trade-off Score Plot(Osc)}
	\label{fig:tradeoff_osc}
	
\end{figure}

As seen from figures \ref{fig:tradeoff} and \ref{fig:tradeoff_osc}, TrustSense held the highest score in all cases where the malicious nodes make up over 20\% of the network. EigenTrust had a significantly lower score than others in all cases due to its heavy resource usage.  The table \ref{tab:ave_score} shows the average score for all reputation systems in order of highest to lowest. 

\begin{table}[h]

	\caption{Average Trade-off Score}
	\setlength{\tabcolsep}{2pt} 
	\small
	\centering
	\def\arraystretch{1.5}
	\begin{tabular}    {| >{\centering\arraybackslash}m{1in} | >{\centering\arraybackslash}m{0.75in} | >{\centering\arraybackslash}m{0.75in} | >{\centering\arraybackslash}m{0.75in} |}
		
		\hline
		\textbf{Reputation System} & \textbf{Static Network} & \textbf{Oscillating Network} & \textbf{Average Overall}\\			
		\hline
		TrustSense & 76.35660327  & 76.44306371 & 76.39983349\\
		\hline
		PeerTrust & 72.58753665 & 68.95800678 & 70.77277172 \\ 		
		\hline
		BTRM & 70.38680544 & 68.41885262 & 69.40282903 \\ 		
		\hline
		PowerTrust & 71.13915581 & 67.25755844 & 69.19835713 \\ 		
		\hline
		EigenTrust & 52.14577244 &	48.92040303 & 50.53308774 \\
		\hline

	\end{tabular}

	\label{tab:ave_score}
\end{table}

\subsection{Discussion}
It could be seen from the above test-results that TrustSense offers a promising advancement in trust/reputation system implementations with respect to resource management and intrusion detection. It opens the potential to producing more miniaturized and longer lasting sensors that will utilize the extra available energy on functional computations with an acceptable level of security for malicious insiders. However, there are still some limitations to the system that confines it applicability in general:

\begin{itemize}
	\item As with other trust/reputation systems, there are topology constraints and TrustSense is not an exception as it can only function in a clustered network.
	\item It might be more expensive getting tamper-proof and long-range cluster-head nodes to facilitate the trust management and evaluation.
	\item TrustSense will not work on highly latent networks like opportunistic systems due to the assumption of reliable and near-real-time communication amongst nodes. 
\end{itemize}	

In the future, we plan to address these limitations in our system in such a way that the balance attained between reliability and energy-efficiency is not skewed.

%% file: conclusion.tex
\section{Conclusion And Future Work}
We conceptualized and prototyped TrustSense, a trust/reputation system for clustered wireless sensor networks that utilizes energy conservation techniques like caching and periodic trust-updates from cluster-heads to attain minimal resource overhead. We also incorporated data anomaly detection using spatial correlation methods, packet loss detection and path estimation. The simulation was carried out using TRMSim-WSN. We compared TrustSense with four other trust reputation systems (PowerTrust, EigenTrust, BTRM and PeerTrust) based on three common metrics which are the accuracy, average path length to a reliable server and average energy consumption. The experimental results show a significant improvement in the energy-to-resiliency trade-off with varying percentages of malicious nodes in the network both in static and oscillating adversary scenarios. We believe this is an additional step in the right direction towards achieving a highly versatile and energy efficient trust reputation scheme for sensor networks.